\documentclass[twocolumn,pra,aps,showpacs]{revtex4}

\usepackage{graphicx}
\usepackage{dcolumn}
\usepackage{amsmath}

\begin{document}

\title{Macroscopic quantum tunneling of 
two-component Bose-Einstein condensates} 
\author{Kenichi Kasamatsu}
\author{Yukinori Yasui}
\author{Makoto Tsubota}
\affiliation{%
Department of Physics,
Osaka City University, Sumiyoshi-Ku,
Osaka 558-8585, Japan \\
}%

\date{\today}

\begin{abstract}
We show theoretically the existence of a metastable state 
and the possibility of decay to the ground state 
through macroscopic quantum tunneling in two-component 
Bose-Einstein condensates with repulsive interactions. 
Numerical analysis of 
the coupled Gross-Pitaevskii equations clarifies the metastable states 
whose configuration preserves or breaks 
the symmetry of the trapping potential, 
depending on the interspecies interaction and the particle number. 
We calculate the tunneling decay rate of the metastable state by 
using the collective coordinate method under the WKB approximation. 
Then the height of the energy barrier 
is estimated by the saddle point solution.
It is found that macroscopic quantum tunneling is 
observable in a wide range of particle numbers. Macroscopic 
quantum coherence between two distinct states is discussed; this 
might give an additional coherent property of 
two-component Bose condensed systems.
Thermal effects on the decay rate are estimated.
\end{abstract}

\maketitle

\section{INTRODUCTION}
\label{intro}
Multicomponent Bose-Einstein condensates (BECs) of alkali-metal atomic 
gases are expected to exhibit macroscopic quantum phenomena 
that have not been found in a single condensate. Multicomponent atomic 
gases can be obtained experimentally by trapping different atomic 
species or the same atoms with different hyperfine spin states. 
The experimental realization of multicomponent 
BECs \cite{Myatt,Stenger,Miesner} further stimulated many 
researchers to study the physics of this interesting system.

Macroscopic quantum tunneling (MQT) is an interesting subject 
in many fields of physics. In this paper we study 
MQT of metastable two-component BECs in a trapping potential. 
Thus we need to know detailed information 
about the stationary state of this system. 
The structure of the ground state has been studied by solving 
two coupled Gross-Pitaevskii equations (GPEs) analytically 
or numerically 
\cite{Ho,Ersy,Pu,Ohberg,Bashkin,Tim,Ersy2,Ao,Tr,Tanatar,Gordon,Ohberg2}.
The stationary solution of the GPEs gives the density profile 
of the condensate characterized by the parameters of the system---trapping 
frequencies, the number of atoms of each component, 
and three s-wave scattering lengths, $a_{1}$, $a_{2}$, and $a_{12}$, 
which represent the interactions between like and unlike components.

The interspecies interaction characterized by $a_{12}$ plays an 
important role in determining the structure of the ground state. 
When the inequality $a_{12}>\sqrt{a_{1}a_{2}}$ is satisfied, a 
mixture of two-component BECs without a trapping potential 
tends to separate spatially \cite{Tim,Ao}. The trapped BECs have two 
different configurations of the condensates when $a_{12}$ is large 
\cite{Ohberg,Ersy2}. One configuration preserves the spatial symmetry 
of the trapping potential by forming a core-shell structure. 
The other breaks the spatial symmetry by displacing the center 
of each condensate from that of the trapping potential. 

Ho and Shenoy first constructed a simple algorithm to determine 
the density profile within the Thomas-Fermi approximation 
(TFA) \cite{Ho}. 
However, the TFA is not enough to describe the density profile of 
phase separation because the penetration at the boundary of 
each component is not considered. Without the TFA 
Pu and Bigelow investigated numerically the ground state of a 
Rb-Na BECs by assuming spherical symmetry \cite{Pu}. 
When $a_{12}$ is large, they found a ground state that forms 
a core of Rb at the center of the trap and a shell of 
Na around Rb, and a metastable state that has a Rb shell 
and Na core. However, they noted the existence of an unstable 
mode which forms the core-shell 
structure. After that, further investigation of two- or 
three-dimensional GPEs showed a spherical symmetry-breaking 
solution for the true ground state \cite{Ohberg,Ersy2,Gordon}.

\"{O}hberg showed that whether the ground state takes 
a symmetry-breaking state (SBS) or a symmetry-preserving 
state (SPS) depends not only on the interspecies interaction 
but also on the particle number, the intraspecies interaction, 
and the shape of the trapping potential \cite{Ohberg2}. However, 
the detaila of the metastable state have not been studied. 
Thus, we investigate the dependence of the ground state 
and the metastable state of two-component BECs in a 
cigar-shaped potential, which can be considered as a quasi-one-dimensional 
system for simplicity. We also make a 
linear stability analysis of the stationary solutions of the 
GPEs and reveal their metastability.

A metastable BEC can also be found in a single condensate 
with negative s-wave scattering 
length \cite{Bradley}. The negative scattering length represents 
an attractive atom-atom interaction, which causes the condensate 
to collapse upon itself to a denser phase. 
The balance between the attractive interaction energy and 
the zero-point kinetic energy of the trapping potential realizes 
the metastable condensate. MQT of a condensate with 
attractive interaction has been predicted \cite{Ueda}. 

For two-component BECs with repulsive interactions 
the metastability mainly comes from the competition between 
intra- and interspecies interactions. We study the 
transition between the SBS and SPS by MQT.

In Sec. \ref{formulation}, we obtain the stationary solution 
of the GPEs numerically. The phase diagram of the ground 
state has a rich structure including metastable states. 
The stability of these solutions is checked by following Ref. 
\cite{Law} which considers 
the stability by taking account of the linear fluctuation 
around the stationary solution. 

In Sec. \ref{MQT}, we introduce the collective coordinate method to 
evaluate the decay rate of a metastable state through MQT 
by an imaginary-time path integral (instanton) technique. 
The collective coordinate enables us to derive an effective 
one-dimensional Lagrangian describing the two-component BECs
and obtain the decay rate.
We estimate the decay rate at finite temperatures; 
the results show the probability of observation 
of MQT. We also discuss macroscopic quantum coherence (MQC), 
which is the oscillation between the SPS and the SBS. 
Section \ref{summary} is devoted to conclusions and discussion.

\section{FORMULATION AND STATIONARY SOLUTION} \label{formulation}
We consider two-component BECs in the external trapping potentials
\begin{equation}
V_{\text{ext}}^{i}({\bf r}) = \frac{1}{2} m_{i} \omega_{i}^{2} x^{2} 
+ \frac{1}{2} m_{i} \omega_{i \perp}^{2} (y^{2}+z^{2}), 
\hspace{3mm}  i=1,2, 
\end{equation}
where $m_{i}$ is the atomic mass, and $\omega_{i}$ 
and $\omega_{i \perp}$ are the longitudinal and transverse 
trapping frequencies. For $\omega_{i \perp} \gg \omega_{i}$ 
the trapping potential is cigar-shaped. If the two-body 
interaction energy is smaller than $\hbar \omega_{\perp}$, 
it does not affect the transverse component $\psi_{i \perp}$ of the 
wave functions, which allows us to analyze the problem in  
one-dimensional space. Although it has been predicted that 
the two-body interaction is changed 
by the effect of tight confinement 
of the trapping potential \cite{Olshanii,Petrov}, 
we will use the following treatment 
to derive the one-dimensional GPEs \cite{Goldstein}.
Using the ground state wave function 
in the harmonic potential for $\psi_{i \perp}(y,z)$, we assume 
the macroscopic wave function as $\Psi_{i}({\bf r} , t) = 
\psi_{i}(x,t) \psi_{i \perp}(y,z) $ $(i=1,2)$. These wave functions 
are substituted into the three-dimensional Gross-Pitaevskii 
energy functional, which is integrated over $y$ and $z$. 
Thus we obtain the one-dimensional Gross-Pitaevskii energy functional,
\begin{eqnarray}
 {\cal H}[\psi_{1},\psi_{2}] &=&  
 \int dx \biggl[  \sum_{i=1,2} 
\biggl( \frac{\hbar^{2}}{2m_{i}} \left| \frac{\partial \psi_{i}}{\partial x} 
\right|^{2} + \frac{1}{2} m_{i} w_{i}^{2} x^{2} |\psi_{i}|^{2}  \nonumber \\ 
& & - \mu_{i} |\psi_{i}|^{2} \biggr) 
 + \frac{1}{2} U_{11} |\psi_{1}|^{4} + \frac{1}{2} U_{22} 
|\psi_{2}|^{4} \nonumber \\
& & + U_{12} |\psi_{1}|^{2}|\psi_{2}|^{2} \biggr] \label{onegpf}
\end{eqnarray}
with the chemical potential $\mu_{i}$. 
Here the two-body interactions $U_{ii}$ and $U_{12}$ are written as
\begin{subequations}
\begin{eqnarray}
U_{ii} &=& g_{ii} \int dy dz \left| 
\psi_{i \perp}(y,z) \right|^{4} = \frac{g_{ii}}{2 \pi b_{i \perp}^{2}},  \\
U_{12} &=& g_{12} \int dy dz \left| 
\psi_{1 \perp}(y,z) \right|^{2} \left| 
\psi_{2 \perp}(y,z) \right|^{2}  \nonumber \\ 
&=& \frac{g_{12}}{\pi (b_{1 \perp}^{2}+b_{2 \perp}^{2})}, 
\end{eqnarray}
\label{interadimension}
\end{subequations}
where $b_{i \perp}=\sqrt{\hbar/m_{i} \omega_{i \perp}}$, 
$g_{ii}=4 \pi \hbar^{2} a_{i}/m_{i}$ and 
$g_{12}=2 \pi \hbar^{2} a_{12}/m_{12}$ with 
reduced mass $m_{12}$. The corresponding two coupled 
time-dependent GPEs are given by
\begin{equation}
i\hbar \frac{\partial \psi_{i}}{\partial t}=
H_{i}[\psi_{1},\psi_{2}]\psi_{i},  \label{2tgpe}
\end{equation}
where
\begin{subequations}
\begin{eqnarray}
H_{1}[\psi_{1},\psi_{2}] &=& -\frac{\hbar^{2}}
{2 m_{1}} \frac{\partial^{2}} 
{\partial x^{2}} + \frac{1}{2} m_{1} \omega_{1}^{2} x^{2} 
- \mu_{1} \nonumber \\
& & + U_{11} |\psi_{1}|^{2} + U_{12} |\psi_{2}|^{2},  \\
H_{2}[\psi_{1},\psi_{2}] &=& -\frac{\hbar^{2}}
{2 m_{2}} \frac{\partial^{2}}
{\partial x^{2}} + \frac{1}{2} m_{2} \omega_{2}^{2} x^{2} 
- \mu_{2} \nonumber \\
& & + U_{22} |\psi_{2}|^{2} + U_{12} |\psi_{1}|^{2},
\end{eqnarray} 
\end{subequations}
and each wave function is normalized by the number 
of particles $N_{i}$ as $\int dx |\psi_{i}(x)|^{2}=N_{i}$.

\subsection{Numerical solution} \label{numerical}
The stationary solutions of Eq. (\ref{2tgpe}) correspond 
to the critical points of the energy functional ${\cal H}$. 
There are several ways to find these critical points numerically. 
Our method is described in the following. The stationary 
solution $\psi_{i0}$ satisfies the relation 
\begin{equation}
H_{i}[\psi_{10},\psi_{20}]\psi_{i0}=0  \label{stationary2gp}
\end{equation}
from Eq. (\ref{2tgpe}). The solution $\psi_{i0}$ 
is taken to be real by making the phase zero.
Using the trial function $\psi_{i}^{\text{tri}}$, 
$\psi_{i0}$ is given by $\psi_{i}^{\text{tri}}$ and its 
deviation $\Delta \psi_{i}$, i.e., 
$\psi_{i0}=\psi_{i}^{\text{tri}}-\Delta \psi_{i}$.
Substituting this relation in Eq. (\ref{stationary2gp}), 
we obtain the linearized equation for $\Delta \psi_{i}$:
\begin{subequations}
\begin{eqnarray}
{H_{1}[\psi_{1}^{\text{tri}},\psi_{2}^{\text{tri}}] + 
2 U_{11} (\psi_{1}^{\text{tri}})^{2}} \Delta \psi_{1} \nonumber \\
+ 2 U_{12} \psi_{1}^{\text{tri}} 
\psi_{2}^{\text{tri}} \Delta \psi_{2} = \sigma_{1}, \\
{H_{2}[\psi_{1}^{\text{tri}},\psi_{2}^{\text{tri}}] + 
2 U_{22} (\psi_{2}^{\text{tri}})^{2}} \Delta \psi_{2} \nonumber \\
+ 2 U_{12} \psi_{1}^{\text{tri}} 
\psi_{2}^{\text{tri}} \Delta \psi_{1} = \sigma_{2},
\end{eqnarray}
\end{subequations}
where $\sigma_{i}=H_{i}[\psi_{1}^{\text{tri}}, \psi_{2}^{\text{tri}}] 
\psi_{i}^{\text{tri}}$. The linear correction 
$\Delta \psi_{i}$ can easily be calculated 
and the modified trial function is defined by
$\psi_{i}=\psi_{i}^{\text{tri}}-\Delta \psi_{i}$.
We repeat the above calculation until the solution converges 
by conserving the norm of each component.

Assuming the condensates of two hyperfine 
spin states of $^{87}$Rb, 
we use the values of the scattering lengths $a_{1}=5.36 $ nm and 
$a_{2}=5.66$ nm, which have the ratio $a_{2}/a_{1}=1.06$ \cite{Matthews}. 
We choose the atomic mass $m_{1}=m_{2}=m_{\text{Rb}}=1.45 \times 
10^{-25}$ kg, the trapping frequency $\omega_{1}=\omega_{2}=\omega=90 
\times 2 \pi$ Hz, and the aspect ratio $\omega_{\perp}/\omega=30$. 
It is convenient to introduce scales characterizing the 
trapping potential: (a) the length scale $b=\sqrt{\hbar/m_{12} \omega}$, 
(b) the time scale $\omega^{-1}$, and (c) the energy scale $\hbar \omega$. 
The dimensionless parameters normalized by these scales 
are expressed by putting a tilde upon 
the symbols.  Then the dimensionless intraspecies interactions 
become $\tilde{U}_{11}=0.2010$ and $\tilde{U}_{22}=0.2123$ from 
Eqs. (\ref{interadimension}). Setting the particle numbers to 
$N_{1}=N_{2}=N$ for simplicity, our formulation has two free parameters, 
$N$ and $\tilde{U}_{12}$. 
The parameter $\tilde{U}_{12}$ might be controlled 
experimentally by the choice of some combination of atoms, or 
by changing the scattering length via the Feshbach resonance \cite{Inouye}.

\begin{figure}[ht]
\includegraphics[height=0.25\textheight]{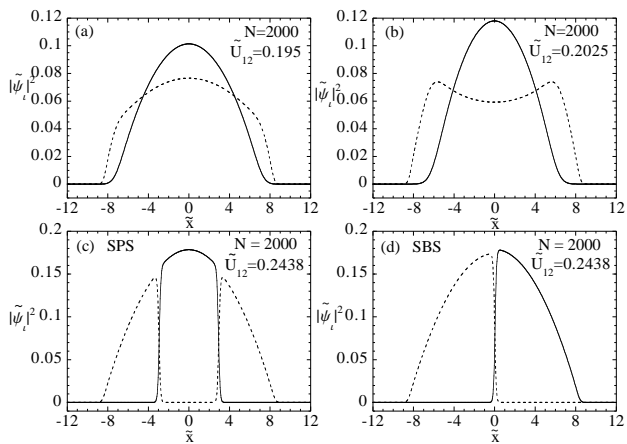}
\caption{Numerical solutions of the two coupled stationary GPEs. 
The solid lines and the dotted lines show the condensates 
1 and 2, respectively. The wave functions $\tilde{\psi}_{i}$ 
divided by $\sqrt{N}$ and the coordinate $\tilde{x}$
are normalized by the length scale $b=\sqrt{\hbar/m_{12}\omega}$. 
(c) shows the symmetry-preserving state with 
energy 87.4261 in units of $\hbar \omega N$.
(d) shows the symmetry-breaking state with 
energy 87.5514.}
\label{stasolfig}
\end{figure}
 
Typical stationary solutions of Eq. (\ref{2tgpe}) are shown 
in Fig. \ref{stasolfig}. When the interspecies interaction is weak, 
two condensates overlap each other. 
For $\tilde{U}_{12} < \tilde{U}_{11} < \tilde{U}_{22}$ 
two overlapping condensates have peaks of the density at the 
center of the trapping potential, as shown in Fig. \ref{stasolfig}(a). 
For $\tilde{U}_{11} < \tilde{U}_{12} < \tilde{U}_{22}$ and 
$ \sqrt{\tilde{U}_{11} \tilde{U}_{22}} < \tilde{U}_{12}$, 
the density peak of condensate 2 is not at 
the center and two peaks appear symmetrically about the origin 
$\tilde{x}=0$, as shown in Fig. \ref{stasolfig}(b). 
Note that the width of condensate 2 is larger than that of 1, 
because $\tilde{U}_{11} < \tilde{U}_{22}$. 
These structures can be predicted easily within the TFA \cite{Ho,Tr}. 
For $\tilde{U}_{12} > \sqrt{\tilde{U}_{11} \tilde{U}_{22}}$ 
the two condensates separate from each other 
with very narrow overlapping regions. 
In Fig. \ref{stasolfig}(c) the condensate  
1 occupies the central region, pushing aside the condensate 
2 symmetrically; this configuration preserves the 
spatial symmetry of the trapping potential. 
On the other hand, as shown in Fig. \ref{stasolfig}(d), 
there exists another configuration with the boundary 
between the two condensates at the center of the trapping 
potential and its spatial symmetry is broken. 
Now we call the stationary state in Fig. 1(c) the 
symmetry-preserving state and that in Fig. \ref{stasolfig}(d) 
the symmetry-breaking state.
The total energy of solution (c) is lower than that of (d) 
as described in the figure caption, so that 
the solution (d) represents the metastable state. 

\begin{figure}[ht]
\includegraphics[height=0.25\textheight]{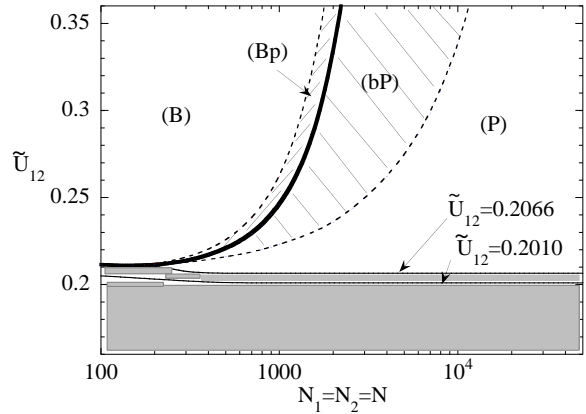}
\caption{$N$-$\tilde{U}_{12}$ phase diagram of the ground state.
The gray region represents the overlapping configuration and 
the other the separated configuration. 
In region $B$ the SBS is the ground state, 
while in region $P$ the SPS is the ground state. 
In the region with slanted lines, there exists  
the metastable SBS (SPS) denoted by the lower-case letter $b$($p$). }
\label{phasedia}
\end{figure}

\begin{figure}[ht]
\includegraphics[height=0.5\textheight]{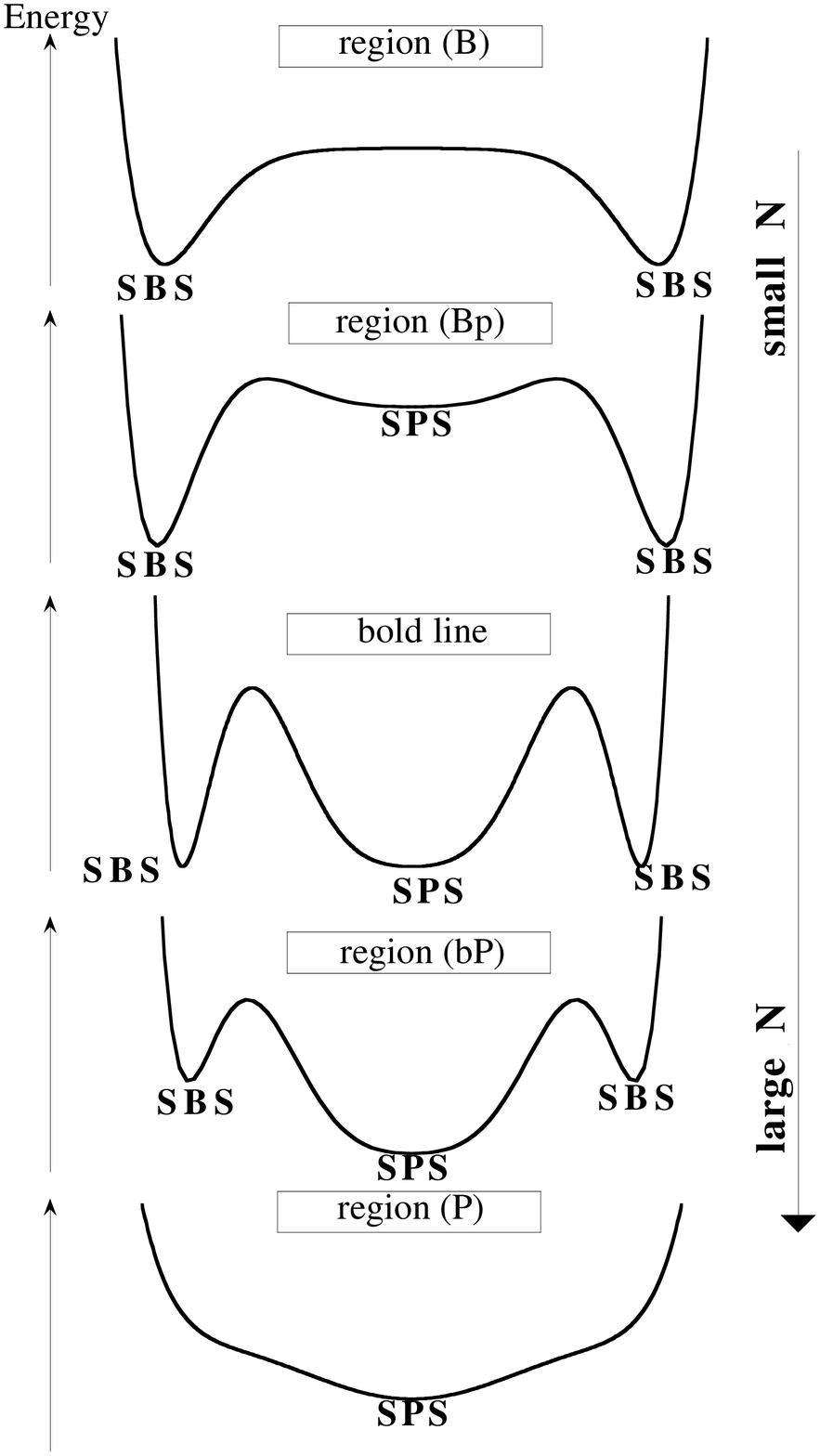}
\caption{Schematic illustration of the total energy in the region 
of the separated configurations in Fig. \ref{phasedia}. 
When the energy of the SBS is equal 
to that of the SPS, the energy configuration becomes a triple well.}
\label{triplewellfig}
\end{figure}

In Fig. \ref{phasedia} we show the $N$-$\tilde{U}_{12}$ 
phase diagram of the ground state. The gray region represents 
the overlapping configurations, Fig. \ref{stasolfig}(a) and \ref{stasolfig}(b), 
and the other the separated configurations, Fig. \ref{stasolfig}(c) 
and \ref{stasolfig}(d); the two regions are divided by the line 
$\tilde{U}_{12}=\sqrt{\tilde{U}_{11} \tilde{U}_{22}} = 0.2066$ 
which was predicted by the TFA 
\cite{Ersy,Bashkin,Tim,Ersy2,Ao,Tr}. 
The gray region is further divided into two regions 
[Fig. \ref{stasolfig}(a) and \ref{stasolfig}(b)] 
by the line $\tilde{U}_{12}=0.2010$ \cite{Tr}. 
More precisely, 
these boundaries are bent for small $N$ 
because of the failure of the TFA.
The region of the separated configurations has the following structure. 
The bold line shows the boundary where the energy of 
the SBS is equal to that of the SPS. In region $B$ 
the SBS is the ground state, 
while in region $P$ the SPS is the ground state. 

The position of the SBS and the SPS in the phase diagram Fig. \ref{phasedia} 
can be understood as follows. 
We first consider the transition on increasing the particle number 
$N$ with a fixed value of $\tilde{U}_{12}$. Note that the SBS 
has one domain wall and the SPS two domain walls. When $N$ is small, 
the SBS is realized because the multiple domain walls increase the 
domain wall energy, which is estimated by the energy $\int dx [\sum_{i} 
(\hbar^{2}/2m_{i}) |\nabla \psi_{i}|^{2} + U_{12} |\psi_{1}|^{2} 
|\psi_{2}|^{2}]$ in Eq. (\ref{onegpf}). The increase in $N$ makes the 
intraspecies interaction energy important, thus tending to extend 
each domain. This overcomes the energy of formation of domain walls, so that 
the SPS becomes more stable than the SBS. When $\tilde{U}_{12}$ increases, 
the domain wall energy becomes larger and thus the region $B$ is extended.

The bold line suggests the existence of metastable states as shown 
in Fig. \ref{triplewellfig}.
In Fig. \ref{phasedia}, 
the regions $Bp$ and $bP$ have metastable states;  
the SBS is the ground state and the SPS is the 
metastable state in the region $Bp$, and vice versa in the region $bP$. 
The details of the metastable state and how to decide 
the boundaries between $B$ and $Bp$, $bP$ and $P$ are 
described in the next subsection.

\subsection{Stability of the solutions} \label{linear}
We linearize the energy functional ${\cal H}$ of Eq. (\ref{onegpf}) 
by substituting
\begin{equation}
\psi_{i}=\psi_{i0} + \delta \psi_{i}.
\end{equation}
Here $\psi_{i0}$ is the stationary solution obtained by solving the  
GPEs, and the fluctuation $\delta \psi_{i}$ is complex. The stationary 
solutions represent the local minima or the saddle points of the 
energy functional. Then the energy can be expanded around the 
stationary solution: 
\begin{eqnarray}
{\cal H}[\psi_{1},\psi_{2}] &\simeq& {\cal H}_{0}+\delta {\cal H}
[\delta \psi_{1},\delta \psi_{2}],  \\
\delta {\cal H} [\delta \psi_{1},\delta \psi_{2}] &=& \frac{1}{2} 
\sum_{i,j=1}^{4} \int d {\bf r} \eta_{i} W_{ij} \eta_{j}.
\end{eqnarray}
Here ${\bf \eta}=(\eta_{1},\eta_{2},\eta_{3},\eta_{4}) \equiv 
(\delta \psi_{1},\delta \psi_{2},\delta \psi_{1}^{\ast},
\delta \psi_{2}^{\ast})$ and $W=(W_{ij})$ is the Hessian operator, 
corresponding to the second-order derivative of the energy functional 
at the stationary solution:
\begin{eqnarray}
W_{11} &=& W_{33}=-\frac{\hbar^{2}}
{2 m_{1}} \frac{\partial^{2}} 
{\partial x^{2}} + \frac{1}{2} m_{1} \omega_{1}^{2} x^{2} 
- \mu_{1}  \nonumber \\
& & \hspace{10mm} +2 U_{11} \psi_{10}^{2} 
+ U_{12} \psi_{20}^{2} , \nonumber \\
W_{22} &=& W_{44}=-\frac{\hbar^{2}}
{2 m_{2}} \frac{\partial^{2}}
{\partial x^{2}} + \frac{1}{2} m_{2} \omega_{2}^{2} x^{2} 
- \mu_{2} \nonumber \\
& & \hspace{10mm} +2 U_{22} \psi_{20}^{2} 
+ U_{12} \psi_{10}^{2} , \nonumber \\
W_{12}&=&W_{23}=W_{34}=W_{41}=U_{12} \psi_{10} \psi_{20},  \nonumber  \\
W_{13}&=&U_{11} \psi_{10}^{2}, \hspace{8mm} W_{24}=U_{22} 
\psi_{20}^{2}.  \nonumber 
\end{eqnarray}
When all eigenvalues of $W$ are positive, the stationary 
solution is stable, while the appearance of a negative 
eigenvalue makes it unstable. 

This eigenvalue problem is simplified 
by the unitary transformation \cite{Law}
\begin{equation}
U^{\dagger} W U = \left(
\begin{array}{ccc}
L_{1} & 0  \\
0 & L_{2} 
\end{array}
\right) ,
\end{equation}
where 
\begin{eqnarray}
L_{1} &=& \left( 
\begin{array}{ccc}
W_{11}-U_{11} \psi_{10}^{2} & 0  \\
0 & W_{22}-U_{22} \psi_{20}^{2}
\end{array}
\right) ,  \\
L_{2} &=& \left( 
\begin{array}{ccc}
W_{11}+U_{11} \psi_{10}^{2} & 2 W_{12}  \\
2 W_{21} & W_{22}+U_{22} \psi_{20}^{2}
\end{array}
\right), \label{2L2} 
\end{eqnarray}
and 
\begin{eqnarray}
U &=& \left( 
\begin{array}{ccc}
{\bf a} & {\bf b}  \\
-{\bf a} & {\bf b}
\end{array}
\right), \nonumber \\
{\bf a} = \left( 
\begin{array}{ccc}
a & 0  \\
0 & a
\end{array}
\right), \hspace{2mm}
{\bf b} &=& \left( 
\begin{array}{ccc}
b & 0  \\
0 & b
\end{array}
\right), \hspace{2mm}
|a|^{2}=|b|^{2}=\frac{1}{2}  \nonumber. 
\end{eqnarray}
Law {\it et al.} used the lowest eigenvalue of $L_{2}$ as 
the stability criterion of the system of two-component BECs 
\cite{Law}. The lowest eigenvalue of $L_{1}$ is zero and the 
eigenfunction is given by the stationary solution $\psi_{i0}$. 

\begin{figure}[bp]
\includegraphics[height=0.42\textheight]{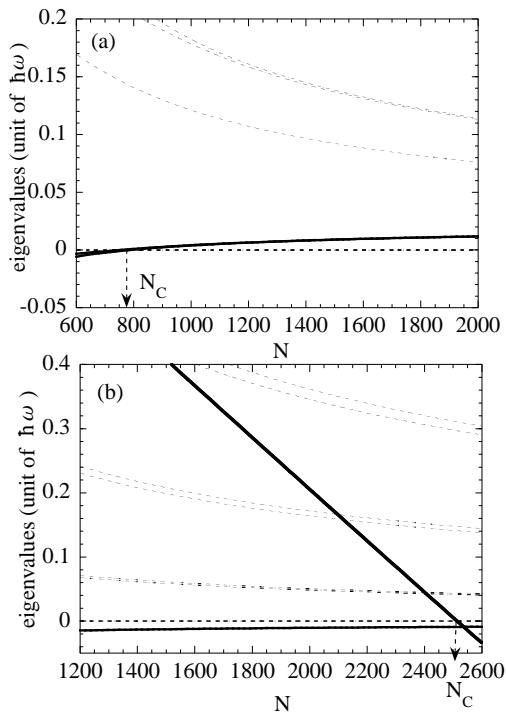}
\caption{Several lower eigenvalues 
of $W$ for the SPS (a) and the SBS (b). 
The bold lines show the eigenvalues of $L_{2}$ 
and the dashed lines show those of $L_{1}$. The critical particle 
number is represented by $N_{c}$.}
\label{eigenfig}
\end{figure}

Figure \ref{eigenfig} shows several lower eigenvalues 
of $W$ as functions of $N$ for the SPS (a) and SBS (b) with 
$\tilde{U}_{12}=0.2438$ as used in Fig. \ref{stasolfig}(c) 
and \ref{stasolfig}(d). 
The critical particle number $N_{c}$ 
defined by the zero eigenvalue of $L_{2}$ gives the criterion 
for the stability of the stationary solution. 
From Fig. \ref{eigenfig}(a), 
the SPS is stable for $N>N_{c}$. Figure \ref{eigenfig}(b) shows 
that there always exists one negative eigenvalue 
whose eigenfunction changes $N$. Hence, as long as $N$ is fixed, 
the SBS is stable for $N<N_{c}$. 
Obtaining $N_{c}$ as a function of $\tilde{U}_{12}$ 
allows us to decide the boundaries between 
the regions $B$ and $Bp$, $P$ and $bP$ in Fig.
\ref{phasedia}. 

Finally, let us note the fluctuation changing the particle number. 
By using the eigenfunction ${\bf u}=(u_{1},u_{2})$ of $L_{2}$ 
this fluctuation is evaluated as
\begin{equation}
\delta N_{i} \simeq 2 \int  \psi_{i0}(x) u_{i}(x) dx.  \label{particlech}
\end{equation}
For the mode in Fig. \ref{eigenfig}(b) whose eigenvalue is always negative 
we obtain $\delta N_{i} \neq 0$. The other mode of $L_{2}$ 
in Fig. \ref{eigenfig}(b) conserves the particle number. 
The fluctuation that changes the 
particle number leads to the ground state of the SBS with 
unbalanced particle number $N_{1} \neq N_{2}$ \cite{Law}. 
 
\section{POSSIBILITY OF MACROSCOPIC QUANTUM TUNNELING} \label{MQT}
As described in Sec. \ref{formulation}, the SBS is the 
ground state and the SPS is the metastable state in 
the region $Bp$ in Fig. \ref{phasedia}, and vice versa 
in the region $bP$. In this section, we study the MQT 
of the metastable state in $Bp$ and $bP$.

\subsection{Collective coordinate approach} \label{collective}
It is difficult to consider 
MQT by full quantum field theory. 
In the case of a single condensate, the variational method 
is often used to estimate the condensate wave function. 
This method was applied to the evaluation of 
the decay rate via MQT of a metastable condensate with 
attractive interaction \cite{Ueda}. 
However, there is an important difference in the description 
of the decay of a single condensate and the 
transition between the SBS and SPS in two-component condensates. 
In the former case, there is an obvious collective 
coordinate, i.e., the spatial size of the condensate 
wave function, which allows us to approximate the wave 
function under the Gaussian ansatz \cite{Ueda}. In contrast, 
in the latter case, it is difficult to find suitable 
collective coordinates that can describe the continuous 
deformation from a metastable state to the ground state.
Thus we introduce an alternative variational approach for this system 
in order to calculate the MQT rate. 

The action $S$ for the Gross-Pitaevskii model is given by 
$S=\int {\cal L} dt$ with the Lagrangian
\begin{equation}
{\cal L} = \sum_{i=1,2} \int dx \left( i \hbar \psi_{i}^{\ast} 
\frac{\partial}{\partial t} \psi_{i} \right) - {\cal H},  \label{action}
\end{equation}
where ${\cal H}$ is the Hamiltonian of Eq. (\ref{onegpf}).
The macroscopic wave functions are written as 
$\psi_{i}=\Phi_{i}(x,t) e^{i \theta_{i}(x,t)}$, 
where $\Phi_{i}^{2}=\rho_{i}$ and $\theta_{i}$ are 
the number density and the phase for each component, 
respectively. Substituting these forms in Eq. (\ref{action}) yields
\begin{equation}
{\cal L} = \int dx 
\left[ \sum_{i=1,2} \left\{ \hbar 
\frac{\partial \rho_{i}}{\partial t} \theta_{i} 
- \frac{\hbar^{2}}{2m_{i}} \Phi_{i}^{2} 
\left( \frac{\partial \theta_{i}}{\partial x} \right)^{2} 
\right\} - V(\Phi_{i}) \right], \label{action2} 
\end{equation}
where $V(\Phi_{i})$ is written as
\begin{eqnarray} 
V(\Phi_{i}) &=& \sum_{i=1,2} \left[ \frac{\hbar^{2}}{2m_{i}} 
\left( \frac{\partial \Phi_{i}}{\partial x} \right)^{2} 
+ (V_{\text{ext}}^{i}-\mu_{i}) 
\Phi_{i}^{2} \right]  \nonumber \\  
& &+\frac{U_{11}}{2} \Phi_{1}^{4} + \frac{U_{22}}{2} 
\Phi_{2}^{4} +U_{12} \Phi_{1}^{2} \Phi_{2}^{2}.  \label{potmujigen}
\end{eqnarray}

The amplitude $\Phi_{i}$ can be expanded around the 
stationary solution $\Phi_{i}^{s}=\sqrt{\rho_{i}^{s}}
=|\psi_{i0}|$ in Sec. \ref{formulation} by using an orthogonal 
complete set $u_{in}(x)$,
\begin{equation}
\Phi_{i}(x,t) = \Phi_{i}^{s}(x) + \sum_{n} \tilde{Q}_{n}(t) 
u_{in}(x) \hspace{5mm} (n=1,2,...),   \label{expden} 
\end{equation}
with the normalization
\begin{equation}
\sum_{i=1,2} \int u_{in} u_{im} dx = \delta_{nm}.
\end{equation}
Here $\tilde{Q}_{n}(t)$ stands for the dimensionless 
arbitrary function and represents the small displacement of 
the density profile from the stationary solution:
\begin{eqnarray}
\sum_{n}\tilde{Q}_{n}^{2}(t)=\sum_{i=1,2} \int dx 
\left[ \Phi_{i}(x,t)-\Phi_{i}^{s}(x) \right]^{2} \\ \nonumber 
(n=1,2,...)  .
\end{eqnarray}
Substituting Eq. (\ref{expden}) to Eq. (\ref{action2}), 
we can obtain the effective action 
written by the functions $\tilde{Q}_{n}(t)$. 
If we assume that the phase has the form $\theta_{i}(x,t)=\sum_{n} 
\tilde{P}_{n}(t) v_{in}(x)/N$ with some complete set $v_{in}(x)$, 
the first term of Eq. (\ref{action2}) can be written as
\begin{equation}
\sum_{nm} 2 \hbar \frac{\tilde{P}_{m}(t)}{N} 
\dot{\tilde{Q}}_{n}(t) \sum_{i} \int dx v_{im}(x) \Phi_{i}^{s}(x) u_{in}(x),
\end{equation}
by using $\dot{\rho}_{i} \simeq 2 \sum_{n} 
\dot{\tilde{Q}}_{n} \Phi_{i}^{s} u_{in}$. 
Then we define the collective 
coordinate $Q_{n}=b\tilde{Q}_{n}/\sqrt{N}$ and the collective 
momentum $P_{m}=2 \hbar \tilde{P}_{m}/b$ with the 
length scale $b$ of the trapping potential. 
By choosing the complete set $v_{in}(x)$ as
\begin{equation}
\sum_{i} \int dx v_{im}(x) \frac{\Phi_{i}^{s}(x)}
{\sqrt{N}} u_{in}(x)=\delta_{mn},
\label{vkime}
\end{equation}
the first term of Eq. (\ref{action2}) becomes 
$\sum_{n} P_{n} \dot{Q}_{n}$ and 
the second term 
\begin{equation}
\int dt \sum_{mn} \frac{P_{m}(t) P_{n}(t)}{2 M_{mn}},
\end{equation}
where the effective mass matrix $M_{mn}$ is given by
\begin{equation}
\frac{1}{M_{mn}}=\sum_{i=1,2} \frac{b^{2}}{4 m_{i} N^{2}} 
\int \rho_{i}^{s} \left( \frac{d v_{im}}{dx} 
\frac{d v_{in}}{dx} \right)  dx.  \label{massmat}
\end{equation}
The potential $V(\Phi_{i})$ is a function of 
the collective coordinate $Q_{n}$. 
Thus we can obtain the effective action
\begin{eqnarray}
S \simeq \int dt \left[ \sum_{n} P_{n} \dot{Q}_{n} - 
{\cal H}_{\text{eff}}({\bf P},{\bf Q}) \right], \label{effaction} \\
{\bf P}=(P_{1},P_{2},...), \hspace{3mm} 
{\bf Q}=(Q_{1},Q_{2},...),  \nonumber
\end{eqnarray}
with the effective Hamiltonian
\begin{equation}
{\cal H}_{\text{eff}}({\bf P},{\bf Q})=\sum_{mn} 
\frac{P_{m} P_{n}}{2 M_{mn}}+V({\bf Q}).
\end{equation}

By substituting Eq. (\ref{expden}) into Eq. (\ref{potmujigen}) 
the effective potential $V({\bf Q})$ can be expanded as follows: 
\begin{equation}
V({\bf Q})=\int V(\Phi_{i}^{s}) dx + \frac{1}{2} \sum_{n} 
\tilde{Q}_{n}^{2} \sum_{i,j} u_{in} H_{ij} u_{jn} + \cdot \cdot \cdot.
\label{finalpot}
\end{equation}
Here the linear term in $Q_{n}$ vanishes because $\Phi_{i}^{s}$ 
is the stationary solution. The quadratic 
term of $Q_{n}$ can be written by the Hessian operator $H_{ij}$, 
which is equal to $2L_{2}$ given by Eq. (\ref{2L2}). 
We take the orthogonal set $u_{in}$ as the eigenfunction of $L_{2}$. 
Thus the second term of 
Eq. (\ref{finalpot}) is diagonalized and $V({\bf Q})$ is written as
\begin{equation}
V({\bf Q})=\int V(\Phi_{i}^{s}) dx+ \frac{1}{2} \sum_{n} N 
\lambda_{n}^{2} \frac{Q_{n}^{2}}{b^{2}} + \cdot \cdot \cdot, \label{potext}
\end{equation}
where $\lambda_{n}^{2}$ $(n=1,2,...)$ are the eigenvalues of $H_{ij}$.
The constant $\int V(\Phi_{i}^{s}) dx$ will be chosen to be zero 
in the following section.

\subsection{Calculation of the decay rate} \label{decayratecal}
We now calculate the MQT rate $\Gamma$ of the 
metastable state in Fig. \ref{phasedia}.
The energy of the metastable state must have an 
(exponentially small) imaginary part 
when the tunneling is taken into account.
Then the decay rate of the metastable 
state is given by \cite{Kleinert}
\begin{equation}
\Gamma = \frac{2}{\hbar} \text{Im} E_{g}.
\end{equation}
The energy $E_{g}$ is evaluated by the 
partition function 
\begin{equation}
Z(\beta) \equiv e^{-W(\beta)} = {\rm tr}(e^{-\beta H}) 
\end{equation} 
as
\begin{equation}
E_{g}=\lim_{\beta \rightarrow \infty} \frac{W(\beta)}{\beta},
\end{equation}
where $\beta=1/k_{B}T$.
Using the action Eq. (\ref{effaction}) with the 
imaginary time $t \rightarrow -i\tau$ (Euclidean action $S_{E}$),
the partition function is written as
\begin{eqnarray}
Z(\beta) = \int D {\bf Q}(\tau) 
\exp \left[ -\frac{S_{E}({\bf Q})}{\hbar} \right]. 
\end{eqnarray}
Within the WKB approximation this path integral 
is evaluated by the saddle-point method \cite{Kleinert}.
More precisely, the dominant contributions to the path integral 
are from paths that minimize the Euclidean action $S_{E}$. 
Such paths are the solution of the Euler-Lagrange equation 
$\delta S_{E}/\delta {\bf Q}=0$, the classical equation of 
motion for the valuables ${\bf Q}(\tau)$ in the inverted 
potential $-V({\bf Q})$. We choose the boundary condition 
that ${\bf Q(\tau)}$ approaches the metastable 
minimum at $\tau=\pm \infty$. By solving this equation 
of motion, we obtain the solution ${\bf Q}_{B}$ 
called the ``bounce solution." 
Then the decay rate has the form
\begin{equation}
\Gamma \simeq A \exp{ \left( -\frac{S_{B}}{\hbar} \right)}, \label{decay}
\end{equation}
where $S_{B}=S_{E}({\bf Q}_{B})$ is the Euclidean action evaluated 
at the bounce solution ${\bf Q}_{B}$ and $A$ the quadratic quantum 
fluctuation around the bounce solution. 
The following describes how to approximate the bounce solution.

We are interested in the regions $Bp$ and $bP$ near the 
dashed lines in Fig. \ref{phasedia}. These regions have 
metastable states as described in Sec. \ref{formulation}. 
On the dashed lines, one of the eigenvalues of the Hessian 
operator $H_{ij}$ vanishes. Thus, in the region close to 
the dashed lines, the potential barrier of the metastable 
state is very small along the direction of the eigenfunction 
with the zero eigenvalue. 
We may assume that the direction of the initial (infinitesimal) 
velocity of the bounce solution is given by the eigenfunction 
${\bf u}_{1}$ subject to the following conditions. First, ${\bf u}_{1}$ 
has the eigenvalue $\lambda_{1}^{2}$ which is small in this 
region and becomes zero on the dashed line. Secondly, this 
eigenfunction conserves the particle number, i.e., 
$\delta N_{i} = 0$ in the sense of Eq. (\ref{particlech}). 
Thus the trajectory of the bounce solution is mainly described 
by the collective coordinate $Q_{1}(t)$, which is the coordinate 
along the direction of ${\bf u}_{1}$, and the other coordinates 
$Q_{2}(t), Q_{3}(t),$ . . . give higher order corrections of the 
solution. These assumptions allow us to solve the bounce solution 
approximately; the trajectory of the bounce solution is straight 
in the collective coordinate space 
${\bf Q}=(Q_{1},Q_{2},Q_{3}$. . . ) \cite{Yasui2}. 
Thus the infinite-dimensional system of Eq. (\ref{effaction}) 
is reduced to a one-dimensional quantum mechanical system 
with the collective coordinate $Q_{1}$ subject to the action
\begin{equation}
S \simeq \int_{- \infty}^{\infty} dt \left[ \frac{M}{2} 
\left( \frac{d Q_{1}}{d t} \right)^{2} + V(Q_{1}) \right].  
\label{omedimaction}
\end{equation}
Here we have defined the mass $M$ by the (1,1) component 
of Eq. (\ref{massmat}); the other components represent 
the mass relevant to $Q_{2},Q_{3}$ . . . , 
so that they are negligible. The mass $M$ includes unknown 
functions $v_{i1}$, although they satisfy Eq. (\ref{vkime}). 
We will assume $v_{i1} \sim  O(1)$, thus obtaining 
$M = M_{11} \sim (4 m_{Rb} N^{2}/b^{2})
\times (b^{2}/N) \sim m_{Rb} N$.

\begin{figure}[bp]
\includegraphics[height=0.25\textheight]{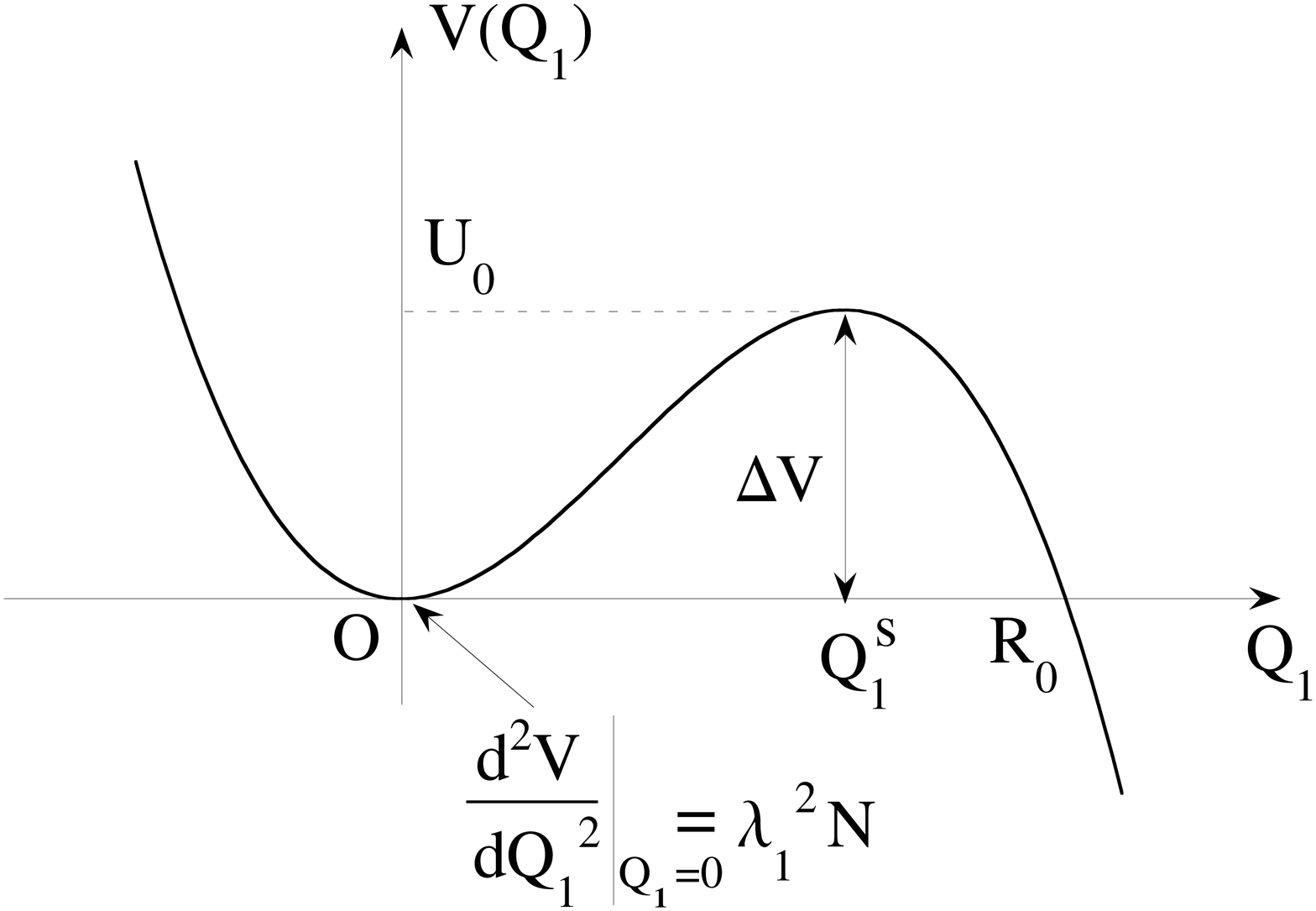}
\caption{The quadratic-plus-cubic potential given by 
Eq. (\ref{apppot}). The potential has a metastable 
minimum at $Q_{1}=0$, barrier height 
$U_{0}=\Delta V$ at $Q_{1}=Q_{1}^{s}$, and 
width $R_{0}$. The second derivative 
of $V(Q_{1})$ at $Q_{1}=0$ is given by $\lambda^{2}_{1} N$.
The coordinate $Q_{1}^{s}$ corresponds 
to the sphaleron with energy $\Delta V$.}
\label{potentialfig}
\end{figure}

\begin{figure}[ht]
\includegraphics[height=0.25\textheight]{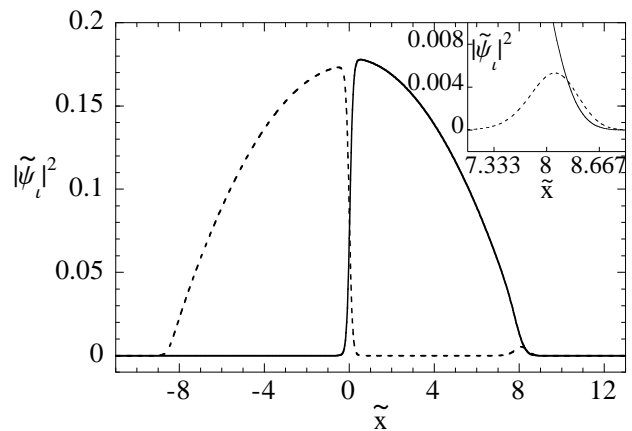}
\caption{Saddle-point solution (sphaleron) 
of the two coupled stationary GPEs 
for $\tilde{U}_{12}=0.2438$ and $N=2000$. 
The unit of length is the same as in Fig.1. 
The inset shows the detail near $\tilde{x}=8$.}
\label{spheraronsol}
\end{figure}

The next problem is to decide the 
form of the effective potential $V(Q_{1})$. 
It should be noticed the potential is expressed as 
$V(Q_{1}) \simeq \frac{1}{2} N \lambda_{1}^{2} (Q_{1}^{2}/b^{2})$ 
for small $Q_{1}$ around the metastable state, from Eq. (\ref{potext}). 
As $Q_{1}$ increases, it is not clear how to extrapolate the potential. 
However, $V(Q_{1})$ should reflect the structure of the original potential 
$V(\Phi_{i})$ of Eq. (\ref{potmujigen}), which has a metastable state in 
addition to the ground state as discussed in Sec. \ref{formulation}. 
Hence we require, first, that $V(Q_{1})$ has a metastable minimum 
at $Q_{1}=0$. Secondly, as $Q_{1}$ increase, $V(Q_{1})$ 
increases once and decreases via a potential barrier $\Delta V$ 
as shown in Fig. \ref{potentialfig}. 
The potential is expected to be written as a power series of $Q_{1}$.
Since the calculation of the decay rate does not need information on 
the ground state, we neglect the $n$th order terms ($n \geq 4$) 
and approximate the potential as 
\begin{equation}
V(Q_{1}) \simeq \frac{1}{2} N \lambda^{2}_{1} \frac{Q_{1}^{2}}{b^{2}} 
- \frac{1}{6} \alpha Q_{1}^{3}  \hspace{3mm} (\alpha > 0), \label{apppot}
\end{equation}
with an unknown parameter $\alpha$. 
Then the height of the potential barrier is given by 
\begin{equation}
\Delta V = \frac{2 N^{3} \lambda_{1}^{6}}{3 \alpha^{2} b^{6}}.
\end{equation} 
The value of $\Delta V$ cannot be determined within the collective 
coordinate method. The barrier $\Delta V$ may be interpreted as follows. 
In general, when we consider quantum tunneling, it is natural to 
assume that the bounce trajectory will go through the sphaleron, 
which is the unstable stationary solution of the equation of motion, 
corresponding to the saddle point of the potential \cite{Manton}. 
Then $\Delta V$ represents the energy of the sphaleron. 
In our case, the equation of motion is the GPE of the 
potential Eq. (\ref{potmujigen}); we have found the sphaleron 
by numerical simulations (see Fig. \ref{spheraronsol}) 
and obtained the value of $\Delta V$. The collective 
coordinate $Q_{1}$ of the sphaleron is simply written as 
\begin{equation}
Q_{1}^{s}=2b \sqrt{\frac{3 \Delta V}{2 \lambda_{1}^{2} N}}.
\end{equation}
Thus our collective coordinate $Q_{1}$ effectively describe 
MQT: the points $Q_{1}=0$ and $Q_{1}=Q_{1}^{s}$ 
correspond to the metastable state and the sphaleron, 
respectively, and the tunneling is represented by the 
bounce solution, going through the sphaleron. 
We will give the explicit bounce solution written via the 
collective coordinate in Eq. (\ref{bouncesol}).

To calculate the decay rate of Eq. (\ref{decay}), 
it is convenient to introduce new scales characterizing 
the quantum tunneling instead of the scales of the 
trapping potential: according to Fig. \ref{potentialfig} we define 
the length scale
\begin{equation}
R_{0} = 3 b \sqrt{\frac{3 \Delta V}
{2 \lambda^{2}_{1} N}},  
\label{chareleng} 
\end{equation}
the energy scale
\begin{equation}
U_{0} = \Delta V, 
\label{charebarrie}
\end{equation}
and a time scale representing the ``tunneling time"
\begin{equation}
\tau_{0} = R_{0} \sqrt{\frac{M}{U_{0}}} = \omega_{0} 
\sqrt{\frac{\hbar}{\omega \lambda^{2}_{1}}} \sqrt{\frac{M}{m_{12} N}}
\label{tunnetimeme}
\end{equation}
with $\omega_{0}=\sqrt{27/2}.$
In these units the action Eq. (\ref{omedimaction}) can be written 
by the dimensionless length 
$q=Q_{1}/R_{0}$ and the time $s=t/\tau_{0}$ as
\begin{eqnarray}
\frac{S}{\hbar} &=& \frac{1}{h} \int_{-\infty}^{\infty} 
ds \left[ \frac{1}{2} \left( \frac{dq}{ds} \right)^{2} 
+\tilde{V}(q) \right], \label{dimlessoneaction}  \\
\tilde{V}(q) &=& \frac{1}{2} \omega_{0}^{2} q^{2}(1-q). 
\label{generalpot}
\end{eqnarray}
Here $h$ is the effective Planck constant defined as
\begin{equation}
h=\frac{\hbar}{\tau_{0} U_{0}}=\frac{1}{\omega_{0}} 
\sqrt{\frac{m_{12} N}{M}} \frac{\sqrt{\hbar \omega 
\lambda^{2}_{1}}}{\Delta V}, \label{mujigenpla}
\end{equation}
whose value must be smaller than unity 
for use of the WKB approximation, 
although it includes the macroscopic valuable $N$. 
From Eq. (\ref{dimlessoneaction}) the bounce solution can 
easily be obtained by solving the equation of motion $d^{2} 
q/d s^{2} = d \tilde{V}(q)/d q$ with the boundary condition $q=0$ 
at $s=\pm \infty$:
\begin{equation}
q_{B}(s)={\text{sech}}
^{2} \left( \frac{\omega_{0}s}{2} \right), \label{bouncesol}
\end{equation}
and the decay rate can be written as
\begin{equation}
\Gamma \simeq  \frac{\tilde{A}}{\tau_{0}} \exp \left( 
-\frac{\tilde{S}_{B}}{h} \right)   \label{findecay}
\end{equation}
with the prefactor $\tilde{A}=4 \sqrt{\omega_{0}^{3}/\pi h} $ 
and the bounce action $\tilde{S}_{B}=8 \omega_{0}/15$.

\begin{figure}[bp]
\includegraphics[height=0.25\textheight]{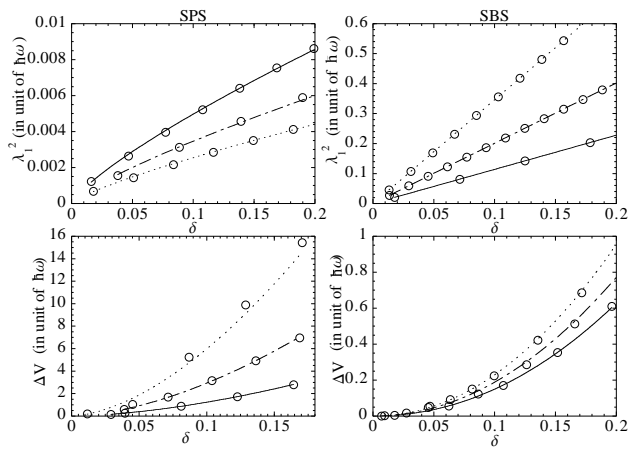}
\caption{$\delta$ dependence of the eigenvalue $\lambda^{2}_{1}$ 
and the potential barrier $\Delta V$. 
The metastable SPS is shown in the left column and 
the metastable SBS in the right column. The open circles represent 
the numerical results of $\lambda_{1}^{2}$ and $\Delta V$ 
from the GPEs. The solid, 
dashed-dotted, and dotted lines show the scaling laws for 
$\tilde{U}_{12}=0.2250, 0.2483,$ and $0.2813$, respectively.}
\label{powerlows}
\end{figure}

We may observe MQT experimentally if the decay rate 
$\Gamma$ is of the order of the lifetime of the BEC. 
Let us search the region near the dashed lines in 
Fig.\ref{phasedia} satisfying this condition.
We obtained the potential barrier $\Delta V$ from the 
sphaleron energy and the eigenvalue $\lambda_{1}^{2}$ 
from the Hessian operator $H_{ij}$ . 
Recalling that $N$ is equal to the critical particle number 
$N_{c}$ on the dashed line (Sec. \ref{linear}), it is 
convenient to introduce a small parameter $\delta=|1-N/N_{c}|$. 
Figure \ref{powerlows} shows the $\delta$ dependence 
of $\lambda_{1}^{2}$ 
and $\Delta V$ for the metastable SPS and SBS near the dashed 
lines in Fig. \ref{phasedia}.
Since $\lambda_{1}^{2}$ and $\Delta V$ vanish on the dashed 
line with $\delta=0$, the scaling laws for the particle number 
are expected to be like those of a single condensate \cite{Ueda}:
\begin{eqnarray}
\lambda^{2}_{1} &\simeq& \hbar \omega 
{\cal C} \delta^{\beta},  \label{power1} \\
\Delta V &\simeq& \hbar \omega {\cal D} \delta^{\gamma}.  
\label{power2}
\end{eqnarray}
The exponents $\beta$,$\gamma$ and the coefficients 
${\cal C}$,${\cal D}$ are determined by 
fitting the scaling laws to the numerical results. 
Thus we obtain the exponents 
$\beta = 1.0 \pm 0.002$, $\gamma = 2.06 \pm 0.03$ for the 
metastable SBS, and 
$\beta = 0.788 \pm 0.006$, $\gamma = 1.673 \pm 0.007$ 
for the metastable SPS. As shown in Table. \ref{ttable}, 
they are approximately independent of the value of $U_{12}$ 
within our analysis, while the coefficients ${\cal C}$ and ${\cal D}$ 
depend on $U_{12}$. When we calculate $R_{0}$ and $U_{0}$ of 
Eqs. (\ref{chareleng}) and (\ref{charebarrie}) by using these exponents 
and coefficients, we find that the metastable SPS has 
larger values of $R_{0}$ and $U_{0}$ than the SBS. 
Thus MQT cannot be expected for the metastable 
SPS compared with the metastable SBS.

Substituting Eq. (\ref{power1}) and 
Eq. (\ref{power2}) to $\tau_{0}$ and $h$ of 
Eqs. (\ref{tunnetimeme}) and (\ref{mujigenpla}), 
we can obtain the scaling 
laws of the decay rate $\Gamma$ of Eq. (\ref{findecay}) by
\begin{equation}
\frac{\tilde{S}_{B}}{h} \simeq \frac{8 \omega_{0}^{2}}{15} 
\sqrt{\frac{M}{m_{12} N}} \frac{{\cal D}}{\sqrt{{\cal C}}} 
\delta^{\gamma-\beta/2},
\end{equation}
and 
\begin{equation}
\frac{\tilde{A}}{\tau_{0}} \simeq \omega \frac{4 \omega_{0}}
{\sqrt{\pi}} \left( \frac{M}{m_{12} N} \right)^{1/4} 
({\cal C} {\cal D}^{2})^{1/4} \delta^
{\beta/4+\gamma/2} . 
\end{equation}

\begin{table}
\caption{The values ${\cal C, D}, \beta, and \gamma$ of 
Eqs. (\ref{power1}) and (\ref{power2}).}

{metastable SPS}

\begin{ruledtabular}
\begin{tabular*}{\hsize}{lccr}
$\tilde{U}_{12}$ & 0.2250 & 0.2483 & 0.2813 \\
\tableline
${\cal C}$ & 0.03041 & 0.021508 & 0.015738  \\
${\cal D}$ & 57.155 & 139.87 & 279.79  \\
\tableline
$\beta$ & 0.78718 & 0.79373 & 0.79177  \\
$\gamma$ & 1.6684 & 1.6801 & 1.671  \\
\end{tabular*}
\end{ruledtabular}

{metastable SBS}

\begin{ruledtabular}
\begin{tabular*}{\hsize}{lccr}
$\tilde{U}_{12}$ & 0.2250 & 0.2483 & 0.2813 \\
\tableline
${\cal C}$ & 1.1445 & 2.0161 & 3.5006  \\
${\cal D}$ & 17.454 & 20.462 & 26.086  \\
\tableline
$\beta$ & 1.0034 & 1.0022 & 1.0055  \\
$\gamma$ & 2.068 & 2.0371 & 2.0562  \\
\end{tabular*}
\end{ruledtabular}
\label{ttable}
\end{table}

\begin{figure}[ht]
\includegraphics[height=0.25\textheight]{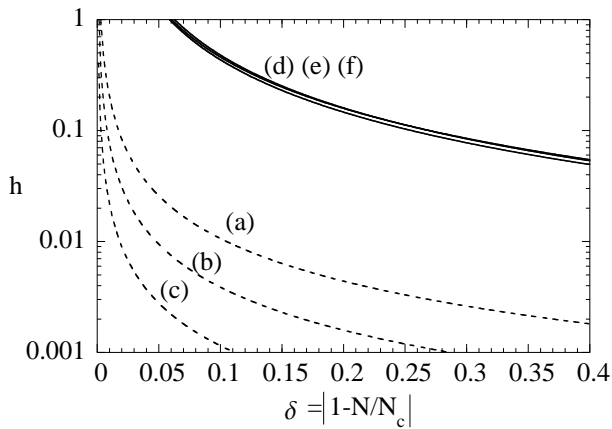}
\caption{The solid lines show the effective Planck 
constant of the metastable SBS and the dashed lines 
show that of the metastable SPS. The parameter 
$\tilde{U}_{12}$ is set as 0.2250 for (a) and (d), 
0.2438 for (b) and (e), and 0.2813 for (c) and (f). 
The critical particle number $N_{c}$ is (a) 481, 
(b) 770, (c) 1197, (d) 1121, (e) 2520, and (f) 5575. 
In the region $h>1$, the WKB approximation breaks down.}
\label{decayratefig}
\end{figure}

The dominant contribution to $\Gamma$
is the exponential factor. To obtain the 
observable decay rate, we require $h \sim 10^{-1}$, although 
$h$ should be small under the WKB approximation.
Figure \ref{decayratefig} shows the effective Planck 
constant for several values of $\tilde{U}_{12}$ as a function of 
$\delta=|1-N/N_{c}|$. The SBS has a wider range with respect 
to $\delta$ satisfying the above condition for $h$ than the SPS. 
Although it is difficult to tune the value of $\delta$ experimentally, 
we may observe MQT of the SBS more easily than that of the SPS.
We now estimate the range of $\delta$ for $U_{12}=0.2483$ 
where $\Gamma$ becomes of the order of $10^{-2}$ sec$^{-1}$; the 
lifetime of the BEC is typically $100$ sec \cite{Myatt}. 
The metastable SPS ($N_{c}=770$) yields $\tilde{S}_{B}/h=9713 
\times \delta^{1.283}$ and $\tilde{A}/\tau_{0}=44.66 \times \omega 
\times \delta^{1.039}$ sec$^{-1}$, so that $\delta < 3.9 
\times 10^{-3}$, a range too narrow to observe MQT. 
For the metastable SBS ($N_{c}=2520$) 
we obtain $\tilde{S}_{B}/h=146.7 \times \delta^{1.537}$ and 
$\tilde{A}/\tau_{0}=53.15 \times \omega \times \delta^{1.269}$ 
sec$^{-1}$; then $\delta < 0.205$. However, values 
of $\delta<6.4 \times 10^{-2}$ make $h$ larger than unity, 
thus breaking the WKB approximation.

\subsection{Macroscopic quantum coherence} \label{mqcdes}
\begin{figure}[bp]
\includegraphics[height=0.25\textheight]{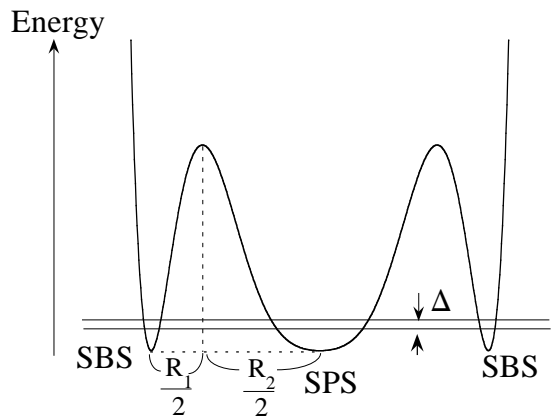}
\caption{Schematic illustration of the triple-well potential.
$\Delta$ shows the splitting of the ground state energy. 
See the text about $R_{1}$ and $R_{2}$.}
\label{mqcpoten}
\end{figure}

An interesting phenomenon may appear on the bold line in 
Fig. \ref{phasedia} where the energy 
of the SPS is equal to that of the SBS. It is 
macroscopic quantum coherence between the SBS and the SPS, 
i.e., the oscillation of a wave packet between their potential wells. 
Then the effective potential $V(Q)$ has a triple-well 
geometry, as illustrated in Fig. \ref{mqcpoten}. 

The period of that oscillation can be estimated by the 
splitting $\Delta$ of the ground state energy due to 
the tunneling. 
The splitting is written as $\Delta \sim A e^{-I/h}$, 
where $h$ is the effective Planck constant, $I$ the instanton 
action, and $A$ a prefactor of the order of unity. 
Here we only estimate the order of the oscillation period by using 
the effective Planck constant $h=\hbar/(R\sqrt{MU_{0}})$ of 
Eq. (\ref{mujigenpla}). 

The barrier height $U_{0}$ is given by the energy of the 
saddle-point solution, and the barrier width $R$ 
is given by the ``distance" between two stable solutions of the 
SBS and the SPS. Then the distance is estimated to 
be $R=(R_{1}+R_{2})/2$ as shown in Fig. \ref{mqcpoten}, where 
$R_{1}$ and $R_{2}$ are given by Eq. (\ref{chareleng}). 
By tuning the parameter $N \sim 200$ and $U_{12} \sim 0.2116$, 
the period of the oscillation becomes of the order of $1$ sec. 
The increase of the particle number raises the energy barrier 
between the SPS and the SBS so that MQC cannot be observed. 

\subsection{Finite-temperature effect} \label{fini}

\begin{figure}[bp]
\includegraphics[height=0.20\textheight]{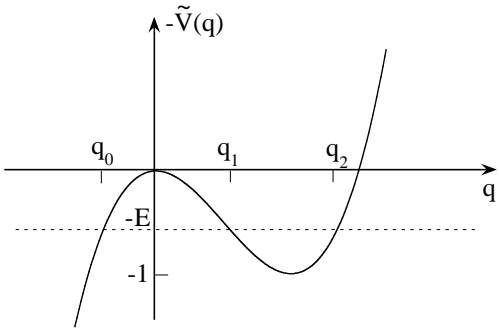}
\caption{Turning points $q_{1}$ and $q_{2}$ in the potential 
$-\tilde{V}(q)=-(1/2)\omega_{0}^{2}q^{2}(1-q)$, 
$\omega_{0}=\sqrt{27/2}$. The ``energy" $E$ ($0<E<1$) 
is determined as a function of inverse temperature 
$\beta$ by requiring that 
the motion between the turning points $q_{1}$ and $q_{2}$ 
is periodic, with period $\beta h$.}
\label{periodicfig}
\end{figure}

Let us consider finite-temperature effects, although 
the discussion of the last subsection was limited to 
zero temperature. Then the bounce solution 
Eq. (\ref{bouncesol}) turns into the periodic solution, 
i.e., the classical solution in the potential 
$-\tilde{V}(q)$ [Eq. (\ref{generalpot})] 
with energy $-E$ $(0<E<1)$ \cite{Zweger}. 
From Fig. \ref{periodicfig} the explicit solution 
is given by the elliptic function
\begin{equation}
q_{B}(s)=q_{2}-(q_{2}-q_{1}) {\text{sn}}^{2} 
\left( \frac{\omega_{0}}{2} \sqrt{q_{2}-q_{0}} s; m \right)
\end{equation}
with $m=\sqrt{(q_{2}-q_{1})/(q_{2}-q_{0})}$, 
and the period $T$ is given by the complete 
elliptic integral of the first kind $K(m)$,
\begin{equation}
T = \frac{4}{\omega_{0}\sqrt{q_{2}-q_{0}}} K(m).
\end{equation}
The period $T$ is related to the temperature, $T=h\beta$, 
where $\beta$ is the inverse temperature normalized by $U_{0}$. 
This solution reduces, of course, to the previous 
solution Eq. (\ref{bouncesol}) for $E=0$. 
The corresponding bounce action is evaluated as
\begin{eqnarray}
\tilde{S}_{B}(\beta) &=& \int_{0}^{h \beta} ds \left[ \frac{1}{2} 
\left( \frac{dq_{B}}{ds} \right) +\tilde{V}(q) \right] \\
     &=& W + h \beta E,
\end{eqnarray}
where
\begin{eqnarray}
W &=&\frac{4}{15} \omega_{0} \sqrt{q_{2}-q_{0}} \nonumber \\
& & \times [ 2 (q_{0}^{2}+q_{1}^{2}+q_{2}^{2}-q_{0} 
q_{1} -q_{0} q_{2} - q_{1} q_{2}) E(m)  \nonumber \\
& & + (q_{1}-q_{0}) (2q_{0}-q_{1}-q_{2}) K(m) ]
\end{eqnarray}
with the complete elliptic integral of the 
second kind $E(m)$. The decay rate by MQT takes the form
\begin{equation}
\Gamma(\beta)=\frac{A(\beta)}{\tau_{0}} \exp 
\left({-\frac{\tilde{S}_{B}(\beta)}{h}} \right) . \label{temp}
\end{equation}
The prefactor $A(\beta)$ was derived in Ref. \cite{Yasui} as 
\begin{eqnarray}
& &A(\beta) = \sqrt{\frac{\omega_{0}^{3}}{2 \pi h}} 
(q_{2}-q_{0})^{3/4} (q_{2}-q_{1}) (1-m^{2}) \nonumber \\
& &\times [a(m)E(m)+b(m)K(m)]^{-1/2} \sinh{\left( 
\frac{\omega_{0} \beta h}{2} \right)}
\end{eqnarray}
with
\begin{eqnarray}
a(m)=2(m^{4}-m^{2}+1) ,  \\
b(m)=(1-m^{2})(m^{2}-1) .
\end{eqnarray}
The thermal effect increases the MQT rate by a factor of only the order 
of unity from the MQT rate at zero temperature.

For $E \rightarrow 0$, we have $(1-m^{2}) 
\sinh(\omega_{0} \beta h /2) \rightarrow 8 $, 
$a(m)E(m)+b(m)K(m) \rightarrow 2$, and $q_{0},q_{1} 
\rightarrow 0$, $q_{2} \rightarrow 1$, so that $A(\beta) 
\rightarrow 4\sqrt{\omega_{0}^{3}/\pi h}$, 
which reproduces the zero-temperature decay rate 
$\Gamma$ of Eq. (\ref{findecay}). 
Let us turn to the limit $E \rightarrow 1$, 
where the period behaves as 
\begin{equation}
\beta h = \frac{2 \pi}{\omega_{0}} \left( 
1+\frac{5}{36}(1-E) + \cdot\cdot\cdot \right).
\end{equation}
The leading term gives the crossover temperature 
$\beta_{c}^{-1}= h \omega_{0}/2 \pi$. 
As the temperature is raised above $\beta_{c}^{-1}$, 
the system has no bounce solutions, and the decay is caused by 
thermal activation (the Arrhenius law): $\Gamma_{\beta} 
\sim \omega_{0} \exp({-\beta \Delta V})$. The crossover temperature 
is of the order of $0.1$nK for the range of 
$\delta$ discussed in the last subsection.

Equation (\ref{temp}) and the Arrhenius formula are not 
available in the narrow region near $\beta_{c}$. 
In this region the decay rate is given by \cite{Affleck}
\begin{eqnarray}
\Gamma(\beta) \tau_{0} &\simeq& \sqrt{\frac{8 \omega_{0}^{3}}
{15 h \pi^{2}}} {\text{sinh}} \left( \frac{\omega_{0} 
\beta h}{2} \right)
{\text{erf}} \left[ \sqrt{\frac{36}{5 \beta_{c}}} 
(\beta-\beta_{c}) \right] \nonumber \\
& & \times \exp \left[ -\beta + \frac{18 \beta_{c}}{5} 
\left( \frac{\beta-\beta_{c}}{\beta_{c}} \right)^{2} \right] 
\label{mediated},
\end{eqnarray}
with the error function 
\begin{equation}
{\text{erf}} (x)=\frac{1}{\sqrt{2 \pi}} 
\int_{- \infty}^{x} dy \exp \left( -\frac{y^{2}}{2} \right) .
\end{equation}
For small $h \ll 10^{-2}$, this formula matches 
smoothly onto Eq. (\ref{temp}) for $\beta>\beta_{c}$ and the Arrhenius 
formula for $\beta<\beta_{c}$ near $\beta_{c}$. However, we cannot apply 
the formula Eq. (\ref{mediated}) to MQT since the value of $h$ 
in our situation is of the order of $10^{-1}$. We leave the 
issue of the crossover region for future study.

\section{CONCLUSIONS AND DISCUSSION} \label{summary}
The metastability and MQT of two-component BECs 
were studied theoretically. 
By analyzing two coupled GPEs numerically, 
we obtained two kinds of metastable state,
the symmetry-breaking state (SBS) and the symmetry-preserving 
states (SPS), which depend on the particle numbers and the
interspecies interaction. We introduced the 
collective coordinate method by improving the usual 
Gaussian variational approach, and calculated the MQT 
rate within the WKB approximation. The effective potential $V(Q)$ was 
determined by analysis of the linear stability 
and using the saddle-point solution. 
Then the decay rate is found to obey a 
scaling law near the critical region. MQT from 
the SBS to the SPS is expected to be observed in a wide 
range of the parameter $\delta$. We also predicted 
MQC between the SBS and the SPS, although the 
range of $\delta$ is rather narrow.

Our analysis is restricted to the 
one-dimensional condensate, but it can be applied 
to a system in a highly anisotropic trapping potential. 
The extension to the three-dimensional 
system is troublesome. However, the 
qualitative nature will be the same as in the one-dimensional case.
This analysis can also be applied to 
the MQT between the domains of a spinor 
condensate \cite{Miesner} where the external magnetic field can be 
used as another variable parameter.

In Sec. \ref{linear} we stated that negative eigenvalues always 
exist for the SBS, corresponding to the change of particle number.
This instability may be caused by inelastic collisions 
of atoms in a real system. However, if we confine ourselves to the region 
near the critical particle number, MQT 
is expected to be the dominant mechanism of decay 
\cite{Ueda}. Thus the change of particle number 
is neglected in the analysis of the MQT. 

Finally, we comment on the validity of the quasi-one-dimensional 
approximation. In this paper, we used the atom-atom interaction 
Eqs. (\ref{interadimension}).
This would be modified for atoms 
in a one-dimensional confining potential 
such as an atom waveguide or a cigar-shaped potential. 
According to Ref. \cite{Olshanii}, 
a two-body potential of the atoms in such a 
confining potential can be written as
\begin{equation}
U_{\rm 1D}(x) = \delta (x) \frac{2 \pi \hbar^{2} a}{m} 
\frac{1}{\pi b_{\perp}^{2}} \left( 1 - 1.4603 \frac{a}
{b_{\perp}} \right)^{-1}, \label{1Dinter}
\end{equation}
where $b_{\perp} = \sqrt{\hbar/m \omega_{\perp}}$. 
For $b_{\perp} \gg a$, which our parameter $b_{\perp} \simeq 
0.29$ $\mu \text{m}$ satisfies, Eq. (\ref{1Dinter}) is smoothly reduced to 
Eq. (\ref{interadimension}).

\end{document}